# Molybdenum Carbide MXenes as Efficient Nanosensors Towards Selected Chemical Warfare Agents

Puspamitra Panigrahi[1*,] Yash Pal [2], Thanayut Kaewmaraya[3, 4], Hyeonhu Bae[5], Noushin Nasiri[6], Tanveer Hussain [7*]

[1] Centre for Clean Energy and Nano Convergence, Hindustan Institute of Technology and Science, Chennai 603103, India

[2] School of Aeronautical Sciences, Hindustan Institute of Technology and Science, Chennai 603103, India.

[3] Department of Physics, Faculty of Science, Khon Kaen University, Khon Kaen 40002, Thailand

[4] Institute of Nanomaterials Research and Innovation for Energy (IN-RIE), NANOTEC KKU RNN on Nanomaterials  Research and Innovation for Energy, Khon Kaen University, Khon Kaen 40002, Thailand

[5] Department of Condensed Matter Physics, Weizmann Institute of Science, Rehovot 7610001, Israel

[6]School of Engineering, Faculty of Science and Engineering, Macquarie University, Sydney, NSW, 2109, Australia

[7]School of Science and Technology, University of New England, Armidale, New South Wales 2351, Australia

Corresponding authors: puspamitrap@hindustanuniv.ac.in, tanveer.hussain@une.edu.au

## Abstract:

There has been budding demand for the fast, reliable, inexpensive, non-invasive, sensitive, and compact sensors with low power consumption in various fields, such as defence, chemical sensing, health care, and safe environment monitoring units. Particularly, an efficient detection of chemical warfare agents (CWAs) is of great importance for the safety and security of the humans. Inspired by this, we explored molybdenum carbide MXenes ($Mo_2CT_x$; $T_x$= O, F, S) as efficient sensors towards selected CWAs, such as arsine ($AsH_3$), mustard gas ($C_4H_8Cl_2S$), cyanogen chloride (NCCl), and phosgene ($COCl_2$) both in aqueous and non-aqueous mediums. Our van der Waals corrected density functional theory (DFT) calculations reveal that the CWAs bind with $Mo_2CF_2$, and $Mo_2CS_2$ monolayers under strong chemisorption with binding energies in the range of -2.33 to -4.05 eV, whereas $Mo_2CO_2$ results in comparatively weak bindings of -0.29 to -0.58 eV. We further report the variations in the electronic properties, electrostatic potentials and work functions of $Mo_2CT_x$ upon the adsorption of CWAs, which authenticate an efficient sensing mechanism. Statistical thermodynamic analysis is applied to explore the sensing properties of $Mo_2CT_x$ at various of temperatures and pressures. We believe that our findings will pave the way to an innovative class of low-cost reusable sensors for the sensitive and selective detection of highly toxic CWAs in air as well as in aqueous media.

## 1. Introduction:

Chemical warfare agents (CWA) are a comprehensive class of toxic volatile organic molecules/compounds (VOCs), which are widely used as chemical weapons due to their high toxicity and lethal effects [1,2]. Since CWAs are cheaply available, these are being easily used as weapons by the groups who are unaware of the consequences [2]. VOCs of CWAs are mostly organic chemicals that have high vapor pressure with a low boiling point (BP) ranges at ordinary room temperature. Due to low BP, most of the VOCs are referred to as gases rather than the liquid compounds and can enter the surrounding environment very easily, thus difficult to monitor. Among common CWAs, arsine ($AsH_3$) is a colorless, flammable, and highly toxic gas with a garlic-like odor which can be detected at concentrations of 0.5 ppm and above. This human carcinogenic $AsH_3$ has been used as a military poison gas in the past. Cyanogen chloride (NCCl) is another organic compound which is a highly toxic blood agent and causes immediate injury upon contact with the eyes or respiratory organs. It can even penetrate the filters in gas masks, and sometimes, it is explosive. Highly poisonous phosgene ($COCl_2$) was used as a chemical weapon during World War I. Exposure to $COCl_2$ causes severe respiratory defects and even death. Similarly, the pale yellow and highly toxic mustard-gas ($C_4H_8Cl_2S$) was also used in World War I. It is corrosive to eyes, skin and lungs and leads to blindness, blistering of skin, and sometimes causes respiratory damages.

These CWAs are generally invisible, inodorous, and at times water soluble, therefore a very accurate and rapid detection is necessary to protect living beings in different environments. For the real-time sensing of the CWAs, the current research emphases on the development of reversible sensing devices which are stable and highly sensitive to detect these CWAs at ppb level with high selectivity both in air and water. In this regard, various sensitive techniques like ion mobility spectrometry [3,4], proton transfer reaction [5], and Raman spectroscopy [6] have been developed to identify different CWAs in the air. However, the above mentioned techniques are mostly delicate, stationary and expensive equipment confined to sophisticated labs and also time-consuming sample preparation are needed to detect CWAs [7]. The current need looks for a practically 'field-able', highly robust, portable, fast, simple-to-operate, and inexpensive detecting method to selectively identify the CWAs in both aqueous and non-aqueous mediums.

Owing to their abundance and excellent sensitivity, many studies have also been conducted on nanostructured metal-oxide-semiconductors (MOS) to develop sensors for CWAs and various VOCs [8–11]. A CuO nanoparticles ZnO flowers shaped heterojunction was reported with high sensitivity towards the detection of certain CWAs [10]. Recently, a portable CWA analyzer was developed with excellent real-time selective detection performance based on Al-doped ZnO quantum dots [12]. Based on their reasonable physicochemical properties and eco-friendly characteristics, $MoO_3$ nanostructures have been used for sensing various VOCs [13,14].

Two-dimensional (2D) nanostructures with a large surface-to-volume ratios, unique optical properties, chemical stability, high electronic conductivity, and ease of functionalization have provided a promising platform in the area of gas sensing [15,16]. Regarding sensitive and selective sensing, carbon nitride ($C_2N$) sheets were investigated as electrochemical sensors for identifying several CWAs [17]. Even various Schottky-contacted dichalcogenides [18,19] and non-carboneous bismuthene type [20] nano-sheets are explored widely to detect various toxic gases but least explored towards sensing VOCs. Recently, various carboneous 2D heterostructures [21] and Pt-decorated phosphorene sheets [22] were investigated for their sensing properties towards various VOCs using first principles simulations[23–26].

MXenes, a unique class of layered 2D materials with general formula $M_{n+1}X_nT_x$, where M are early transition metals, X is C or N and $T_x$ presents surface termination groups like -H, -OH,-S, -F or –Cl, has been hotly pursued for various applications [27]. MXenes with enhanced mechanical flexibility, stretchability, and ionic conductivity have a wide range of applications in wearable sensors [28]. Ti-, and Mo-based MXenes ($Ti_2CT_x$, $Mo_2CT_x$) with rich, localized states near the Fermi level have been studied for the sensing of toxic gases [29–31]. However, one of the vital issues for most of the gas sensors is their sensing performances in the presence of water [28], as the detection of a target gas is very different in humid air than in dry conditions [32]. Since the gas sensing process is strongly related to surface reactions, the sensitivity of MXenes will also change with the factors influencing the surface reactions, like the microstructures of the sensing layers with water [33,34]. Other than that, an oriented electric field can also affect the outcomes of the electronic interactions in such a way that the sensing properties and adsorption strength between the target gaseous molecules and the sensor materials get altered [35]. It is possible to modify the rate and selectivity of a catalytic reaction by using an external electric field, as long as the orientation of the electric field is aligned in favor of electron flow [36]. Various 2D materials showed outstanding sensing aptitudes under external electric field [37,38].

Here, using first principles simulations, we investigated the sensing propensity of $Mo_2CT_x$ ($T_x$: O, S, F) towards highly toxic CWAs like arsine ($AsH_3$), mustard gas ($C_4H_8Cl_2S$), cyanogen chloride (NCCl), and phosgene ($COCl_2$) both in aqueous and non-aqueous mediums. We further explored the roles of various surface functionals groups on the sensing mechanism of $Mo_2CT_x$.

## 2. Methodology

All the first principles density functional theory (DFT) calculations were executed by using the Vienna ab-initio Simulation Package (VASP) [39]. The projector-augmented-wave (PAW) method [40], in combination with the generalized gradient approximation (GGA) and Perdew-Burke-Ernzerhof (PBE) [39,41–43] were used to handle the exchange-correlation functional. The Kohn-Sham orbitals were expanded with a plane-wave basis set having a converged cut-off energy of 500 eV. The proposed three layered $Mo_2CT_x$ monolayers were modelled by taking 3 x 3 x1 supercell while maintaining a vacuum of 20 Å along perpendicular direction to avoid interaction between the periodic images. To obtain the spin-polarized ground state structures and reveal the electronic properties, the calculations were performed within Brillouin zone sampling with a (3 × 3 × 1) and a much denser ( 5 x 5 x1) k-point mesh respectively within Monkhorst-Pack scheme [44]. All the structures were entirely loosened until all of the unconstrained atomic force on each ion was converged to 0.001 eV Å$^{-1}$ by using the quasi-Newton algorithm. The energy convergence of 10$^{-6}$ eV was specified throughout the self-consistent calculations. For accurate estimation of the binding energies of non-covalently bonded CWAs to $Mo_2CT_x$, the van der Waals (vdW) dispersion corrected DFT-D3 approach proposed by Grimme[43] were implemented to the entire calculations.

The Bader charge analysis [45,46] is performed to quantify the loss or gain of charge clouds during the adsorption of CWAs to the proposed $Mo_2CT_x$.

The adsorption energies of CWAs to $Mo_2CT_x$ are calculated as,

$$E_{ads} = [E(Mo_2CT_x@CWAs) - E(Mo_2CT_x) - E(CWAs)] \quad\quad\quad\quad\quad (1)$$

Here, first, second, and third terms represent the total energies of $Mo_2CT_x$ adsorbed with CWAs, bare $Mo_2CT_x$, and isolated CWAs molecules, respectively.

Further, to apply the static external electric field, the method projected by Neugebauer and Scheffler [47], as implemented in VASP, was used. The idea is to generate a uniform electric field along the perpendicular Z-direction. For this purpose, an adequate vacuum space needed along the Z-direction to avoid the overlapping of the charge density between the $Mo_2CT_x$ and the artificial dipole sheet. Conversely, with a wider vacuum thickness, the electrons near the Fermi surface will instead get pulled away by the strong electric field, resulting in field emission [36,47]. As reported earlier, an electron can tunnel to the vacuum when the distance between the surface and the artificial dipole sheet exceeds $\phi/E_f$ where $\phi$ and $E_f$ are the work function and applied electric field, respectively.

### 3. Results and Discussion:

#### 3.1 Structural and Electronic properties of $Mo_2CT_x$

In the first phase, we will discuss the structural and electronic properties of $Mo_2CT_x$. Ground state structures and their corresponding spin-polarized density of states (SPDOS) are presented in Fig. 1(a-f). The SPDOS plots with densely populated states near the Fermi level and revealing the symmetry in spin-up and spin-down states, predict the metallic as well as non-magnetic behavior of the $Mo_2CT_x$ monolayers (Fig 1. d-f).

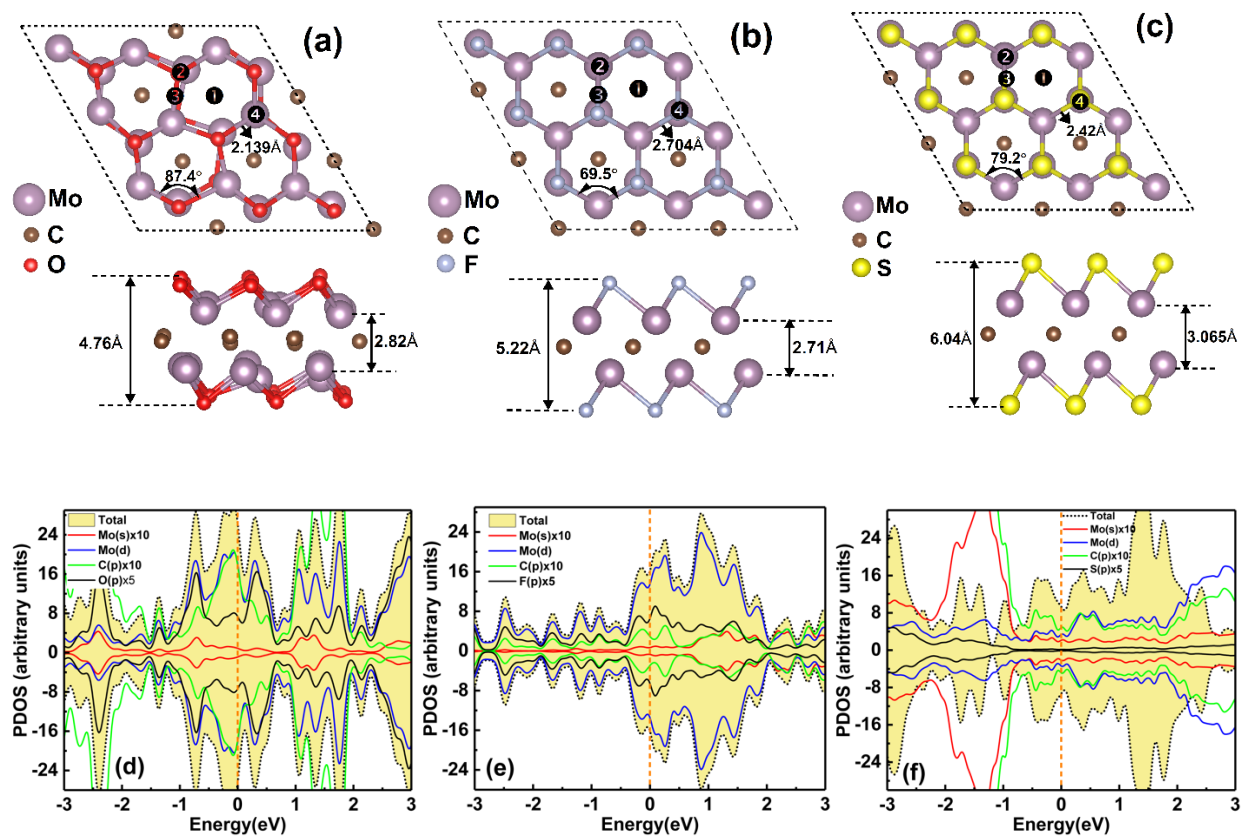

Fig. 1: Top and side views of (a)$Mo_2CO_2$, (b) $Mo_2CF_2$, (c) $Mo_2CS_2$ labeled with bond lengths, buckling parameter, bond angles, and (d-f) their corresponding spin-polarized partial density of states (PDOS), respectively. The dotted line represents the Fermi level.

### 3.2 Adsorption of CWAs on $Mo_2CT_x$

As mentioned, we have identified four crucial CWAs, $AsH_3$, $C_4H_8Cl_2S$, NCCl, and $COCl_2$, to be selectively detected by $Mo_2CT_x$. To reveal the exact adsorption sites of selected CWAs on the $Mo_2CT_x$, different binding sites (site 1, 2, 3, 4 in Fig. 1a-c), have been explored. Furthermore, CWAs are exposed to each monolayer with three possible orientations. For example, to identify the best adsorption orientation of $AsH_3$, both H and As site were exposed to the $Mo_2CT_x$ separately. The adsorption energy of the studied CWAs on $Mo_2CT_x$ at all possible orientations are summarized in Table S1 (Supporting Information). The exothermic process identifying the best possible adsorption configurations of CWAs adsorbed $Mo_2CT_x$ ($Mo_2CT_x$@CWAs) are summarized in Table 1 and are considered for the further investigations. The ground state structures and the corresponding SPDOS of $Mo_2CT_x$@$AsH_3$, $Mo_2CT_x$@$C_4H_8Cl_2S$, $Mo_2CT_x$@NCCl, and

Mo$_2$CT$_x$@COCl$_2$ are presented in Figs. 2, 3, 4(a-c). The corresponding binding energies and equilibrium distances are given in Table 1.

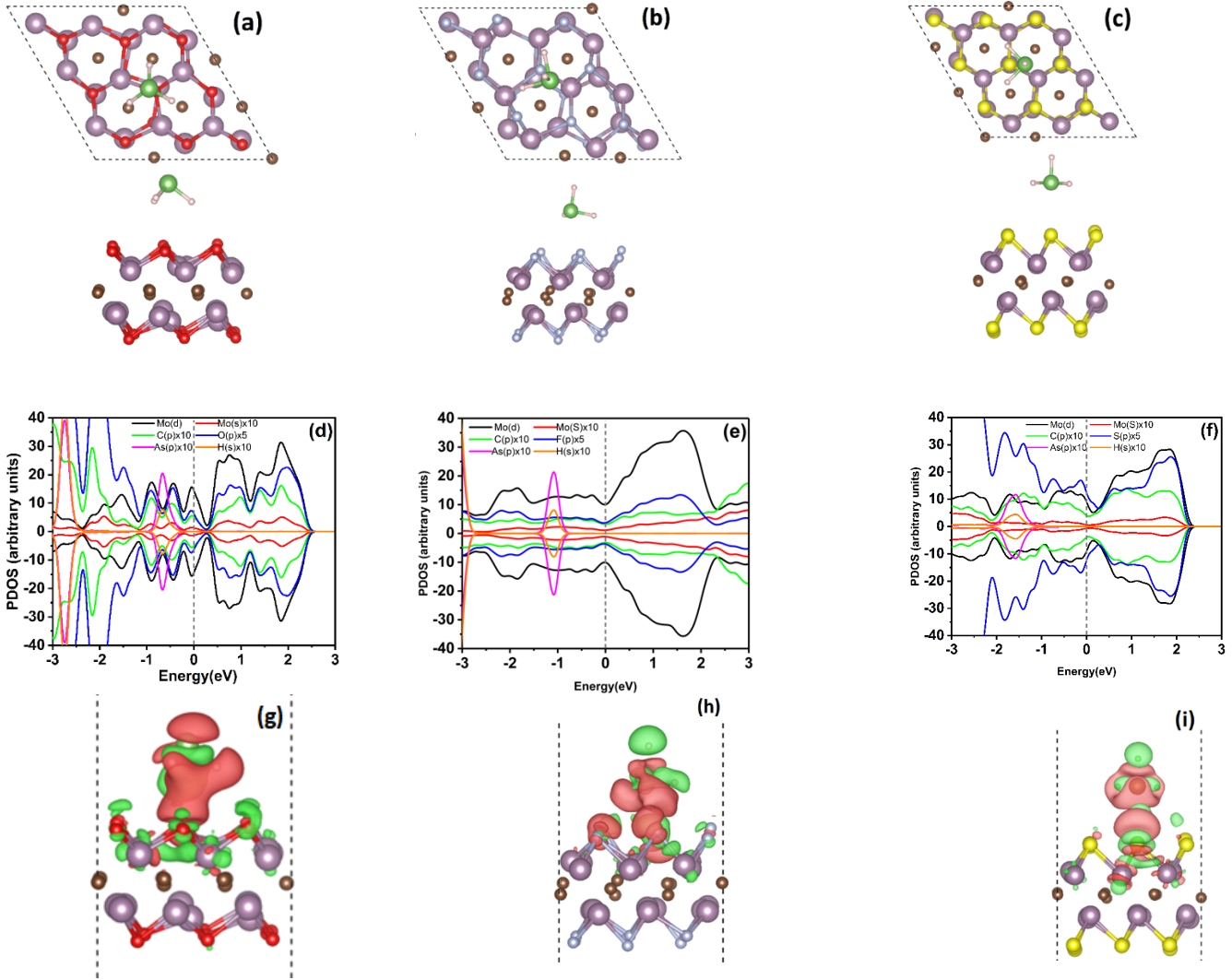

Fig. 2: (a-c) top and side views of the optimized structures, (d-f) the corresponding partial density of states, and (g-i) iso-surface charge density plots of AsH$_3$ adsorbed Mo$_2$CO$_2$, Mo$_2$CF$_2$, and Mo$_2$CS$_2$, respectively. (Red: charge depletion and Green: charge accumulation with iso-value =0.001 Bohr$^{-3}$).

Table 1. The binding distance (Δd), adsorption energies (E$_{ads}$), and net charge transfer (Δρ) of Mo$_2$CT$_x$@CWAs. A positive (negative) indicates a loss (gain) of electrons from each CWAs to the Mo$_2$CT$_x$ monolayers.

| CWAs | Mo$_2$CT$_x$@CWAs | | | | | | | | |
|---|---|---|---|---|---|---|---|---|---|
| | Mo$_2$CO$_2$ | | | Mo$_2$CF$_2$ | | | Mo$_2$CS$_2$ | | |
| | Δd (Å) | $E_{ads}$ (eV) | Δρ (e) | Δd (Å) | $E_{ads}$ (eV) | Δρ (e) | Δd (Å) | $E_{ads}$ (eV) | Δρ (e) |
| AsH$_3$ | 2.86 | -0.29 | -2.67 | 2.85 | -4.05 | -2.41 | 3.10 | -1.98 | -2.11 |
| C$_4$H$_8$Cl$_2$S | 2.66 | -0.38 | -2.40 | 1.48 | -3.07 | -2.33 | 2.90 | -2.54 | -2.20 |
| NCCl | 2.92 | -0.58 | -2.58 | 2.83 | -3.93 | -2.77 | 2.89 | -2.02 | -2.82 |
| COCl$_2$ | 2.81 | -0.40 | -2.37 | 2.07 | -4.12 | -2.48 | 2.25 | -1.99 | -2.38 |

It is evident from Fig. 2 (a-c) that AsH$_3$ stabilizes itself with its 'H' atoms pointing towards Mo$_2$CO$_2$, however tilted configurations are observed on Mo$_2$CF$_2$ and Mo$_2$CS$_2$ with 'A' atom pointing towards the sheets. The AsH$_3$ binds at optimized distances of 2.86, 2.85 and, 3.10 Å on Mo$_2$CO$_2$, Mo$_2$CF$_2$, and Mo$_2$CS$_2$, respectively. The SPDOS plot of Mo$_2$CO$_2$@AsH$_3$ (Fig. 2d) depicts a strong hybridization between C-2p and O-2p with As-4p state. In Mo$_2$CF$_2$@AsH$_3$ (Fig. 2e), hybridization between H-1s with F-2p/S-3p states are evident near the Fermi surface. In case of Mo$_2$CS$_2$@AsH$_3$, within -0.1 eV to the edge of Fermi $E_f$ in the valence band region the majority of contribution arises from the AsH$_3$ whereas a majority of Mo-4d bands gets shifted towards the conduction band region as shown in Fig. 2f.

Fig. 3 (a-c) represents the lowest energy configurations of Mo$_2$CT$_x$@C$_4$H$_8$Cl$_2$S. Here the 'H' atoms of the C$_4$H$_8$Cl$_2$S point towards the Mo$_2$CT$_x$ monolayers. The C$_4$H$_8$Cl$_2$S binds at optimized distances of 2.66, 1.48 and, 2.90 Å on Mo$_2$CO$_2$, Mo$_2$CF$_2$, and Mo$_2$CS$_2$, respectively. The SPDOS plot of Mo$_2$CO$_2$@C$_4$H$_8$Cl$_2$S reveals H-*1s* and Cl-*3p* states of C$_4$H$_8$Cl$_2$S hybridize with O-*2p* and C-*2p* states of Mo$_2$CO$_2$. In Mo$_2$CF$_2$@C$_4$H$_8$Cl$_2$S, the F-*2p* and C-*2p* states of the Mo$_2$CF$_2$ hybridizes with S-3p states of the C$_4$H$_8$Cl$_2$S. Similarly, the C-2p and S-3p states of the Mo$_2$CS$_2$ can been seen hybridizing with Cl-3p and H-1s of the C$_4$H$_8$Cl$_2$S in case of Mo$_2$CS$_2$@C$_4$H$_8$Cl$_2$S. It is worth mentioning that the valence band region within -0.05 eV to the edge of Fermi (E$_f$) reveals sharp peak with contribution from S-3p and H-1s being hybridized with Mo-4d states in Mo$_2$CO$_2$@C$_4$H$_8$Cl$_2$S. For Mo$_2$CF$_2$@C$_4$H$_8$Cl$_2$S, both C-2p and Cl-3p states get shifted towards the conduction band region. In $Mo_2CS_2@C_4H_8Cl_2S$, the valence band region within -0.08 eV to the edge of $E_f$ are dominated with states from the $C_4H_8Cl_2S$ being hybridized with Mo-4d states.

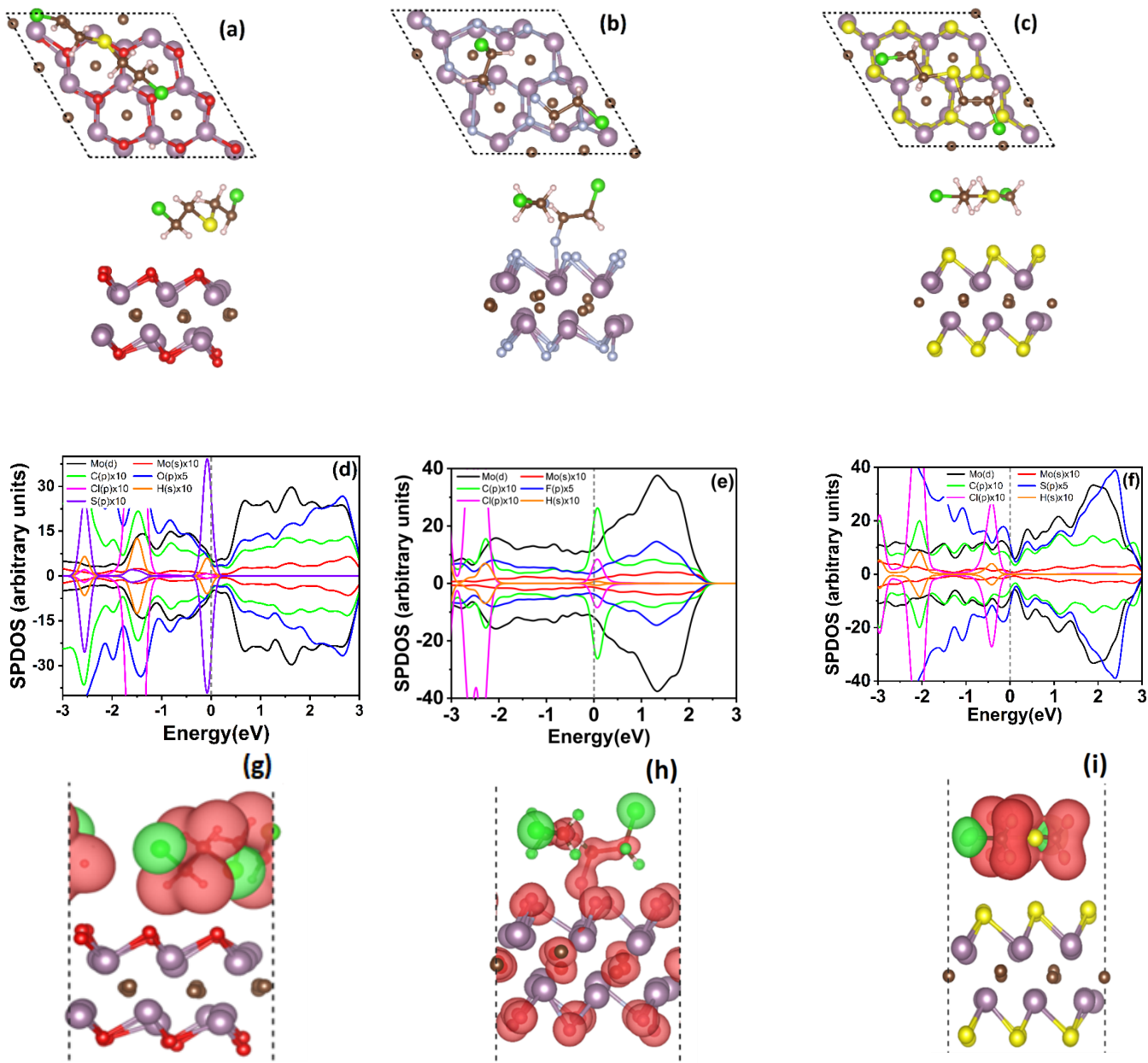

Fig.3: (a-c) top and side views of the optimized structures, (d-f) the corresponding partial density of states, and (g-i) iso-surface charge density plots of $C_4H_8Cl_2S$ adsorbed $Mo_2CO_2$, $Mo_2CF_2$, and $Mo_2CS_2$, respectively. (Red: charge depletion and Green: charge accumulation with iso-value =0.001 Bohr$^{-3}$).

The NCCl prefers a horizontal geometry over $Mo_2CO_2$, however H-directed configurations have been observed over $Mo_2CF_2$, and $Mo_2CS_2$ monolayers, as shown in Fig. 4 (a-c). The NCCl binds at optimized distances of 2.92, 2.83 and, 2.89 Å on $Mo_2CO_2$, $Mo_2CF_2$, and $Mo_2CS_2$, respectively.

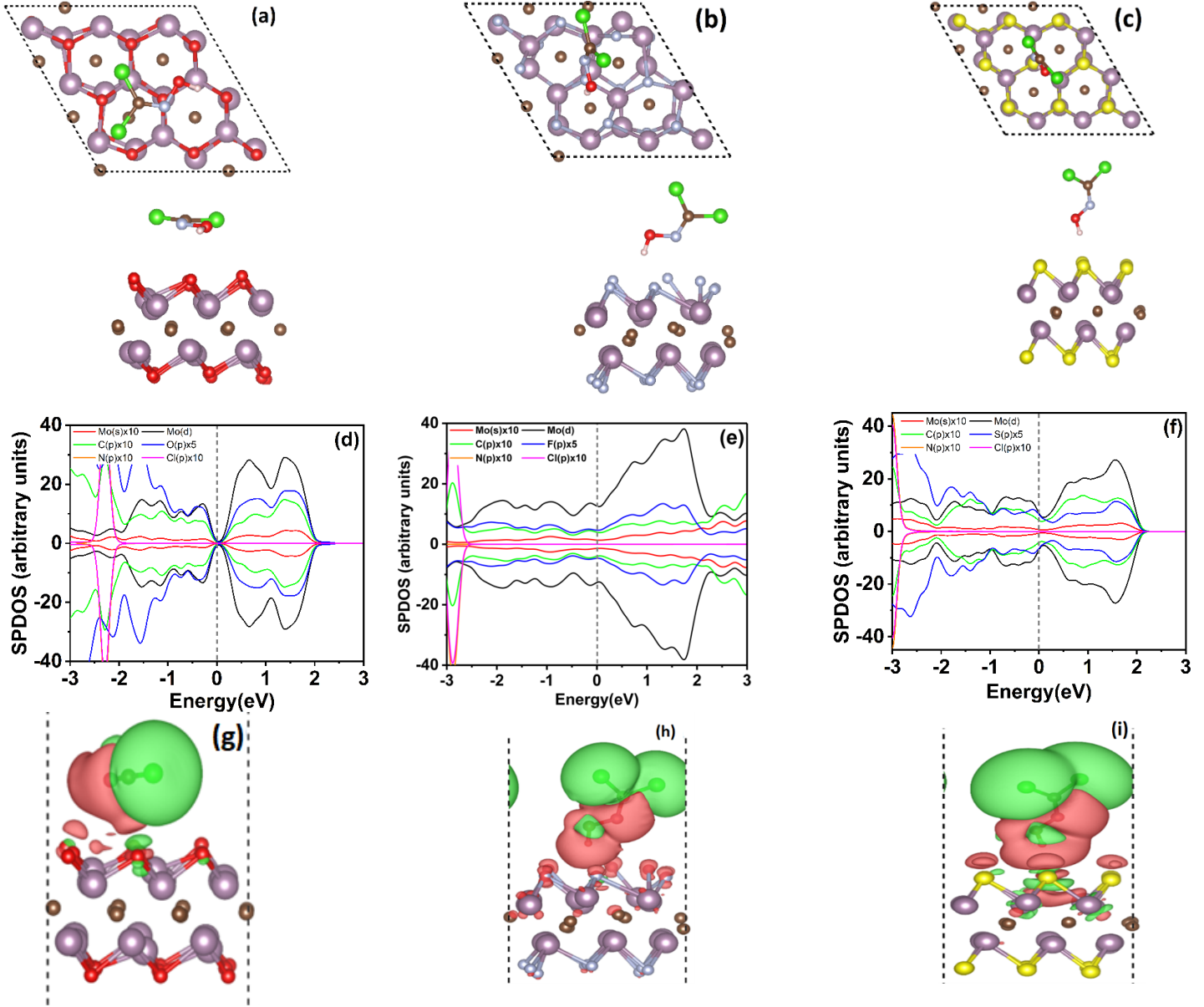

Fig.4: (a-c) top and side views of the optimized structures, (d-f) the corresponding partial density of states, and (g-i) iso-surface charge density plots of NCCl adsorbed $Mo_2CO_2$, $Mo_2CF_2$, and $Mo_2CS_2$, respectively. (Red: charge depletion and Green: charge accumulation with iso-value =0.001 Bohr $^{-3}$).

The adsorption of NCCl brings significant distortion to the electronic structures of $Mo_2CT_x$ as compared to $AsH_3$ and $C_4H_8Cl_2S$. In the case of $Mo_2CO_2$@NCCl (Fig. 4d), SPDOS plot reveals a sharp valley like features near the $E_f$ region. Top edge of valance band near $E_f$ Mo-*4d* states hybridizes with O-*2p* states whereas conduction band tailing is dominated with Mo-4d empty states. For both $Mo_2CF_2$@NCCl, and $Mo_2CS_2$@NCCl, Fig. 4 (e, f), the Mo-*4d* states gets shifted towards conduction band region revealing significant charge transfer whereas, valance bands are dominated with adsorbate states being hybridized with the host sheet.

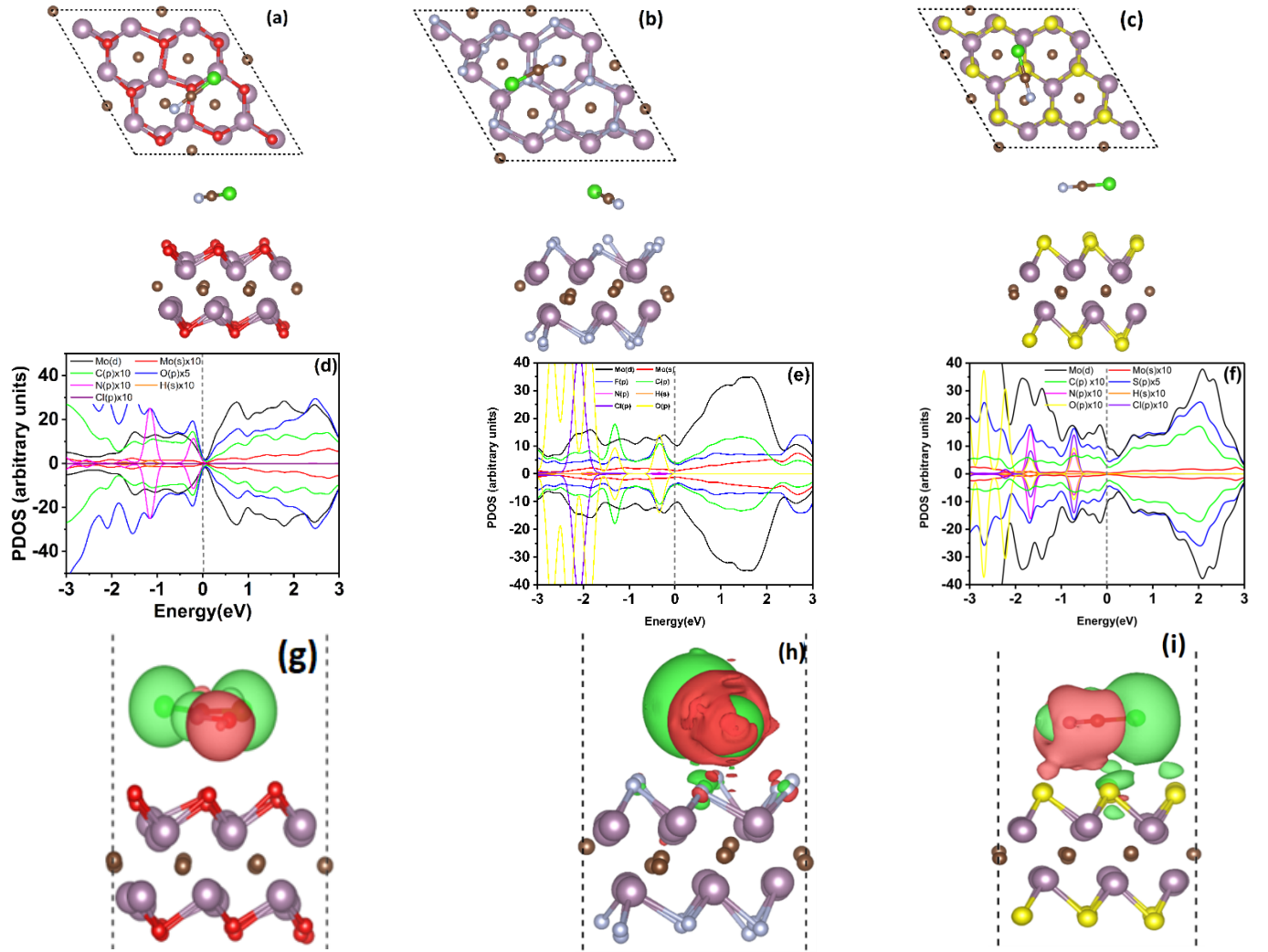

Fig.5: (a-c) top and side views of the optimized structures, (d-f) the corresponding partial density of states, and (g-i) iso-surface charge density plots of $COCl_2$ adsorbed $Mo_2CO_2$, $Mo_2CF_2$, and Mo$_2$CS$_2$, respectively. (Red: charge depletion and Green: charge accumulation with iso-value =0.001 Bohr$^{-3}$).

The ground state geometries of Mo$_2$CT$_x$@COCl$_2$ show almost horizontal alignment of the COCl$_2$ over the studied monolayers, as shown in Fig. 5 (a-c). The COCl$_2$ binds at optimized distances of 2.81, 2.07 and, 2.25 Å on Mo$_2$CO$_2$, Mo$_2$CF$_2$, and Mo$_2$CS$_2$, respectively. The corresponding SPDOS plots, Fig.5 (d-f), indicate a strong hybridization between the COCl$_2$ and Mo$_2$CT$_x$, which further manifests the strong orbital interaction between the N-*2p* and C-*2p* and O-*2p* states verifying their strong E$_{ads}$ values. The binding mechanism of CWAs on Mo$_2$CT$_x$ has been further explained by Bader charge analysis. Charge analysis reveals that in addition to the transfer of charges among the CWAs and Mo$_2$CT$_x$, there is exchanges of charges within the Mo$_2$CT$_x$ monolayers. The charge transfer of all the studied systems are summarized in Table 1 and corresponding charge redistribution plots are depicted Fig. 2,3,4 and 5 (g-i). In comparison to Mo$_2$CO$_2$, the substantial charge redistribution explains the strong adsorption of CWAs on Mo$_2$CS$_2$ and Mo$_2$CF$_2$ monolayers.

## 3. 3 Electrostatic potential and work function

Upon the adsorption of CWAs, the charge redistribution in Mo$_2$CT$_x$ is distressed indicating a distinct variation in resistances of the monolayers. Since the change in resistance of a sensing materials (Mo$_2$CT$_x$ here) is an essential parameter for efficient sensing mechanism, therefore it is important to calculate the planer average of the electrostatic potentials (Vz) [19,20]. The Vz values of Mo$_2$CT$_x$ and Mo$_2$CT$_x$@CWAs as per the following eq. (2), and the results are presented in Fig.6 (a-c).

$$\bar{V}(z) = \frac{1}{A} \iint_{cell} V(x,y,z) \, dxdy \qquad (2)$$

Here, A is the area of the modelled Mo$_2$CT$_x$. While plotting $\bar{V}(z)$ as a function of *z*, one can extract $V(\infty)$, the electrostatic potential in vacuum. It is interesting to see that the Vz of Mo$_2$CT$_x$@CWAs systems gets distinctively enhanced upon the adsorption of each CWAs near the Fermi as compared to the Vz of the bare $Mo_2CT_x$. The enhanced out-of-plane electric field in the case of $Mo_2CF_2$@CWA explains the strong adsorptions of CWAs.

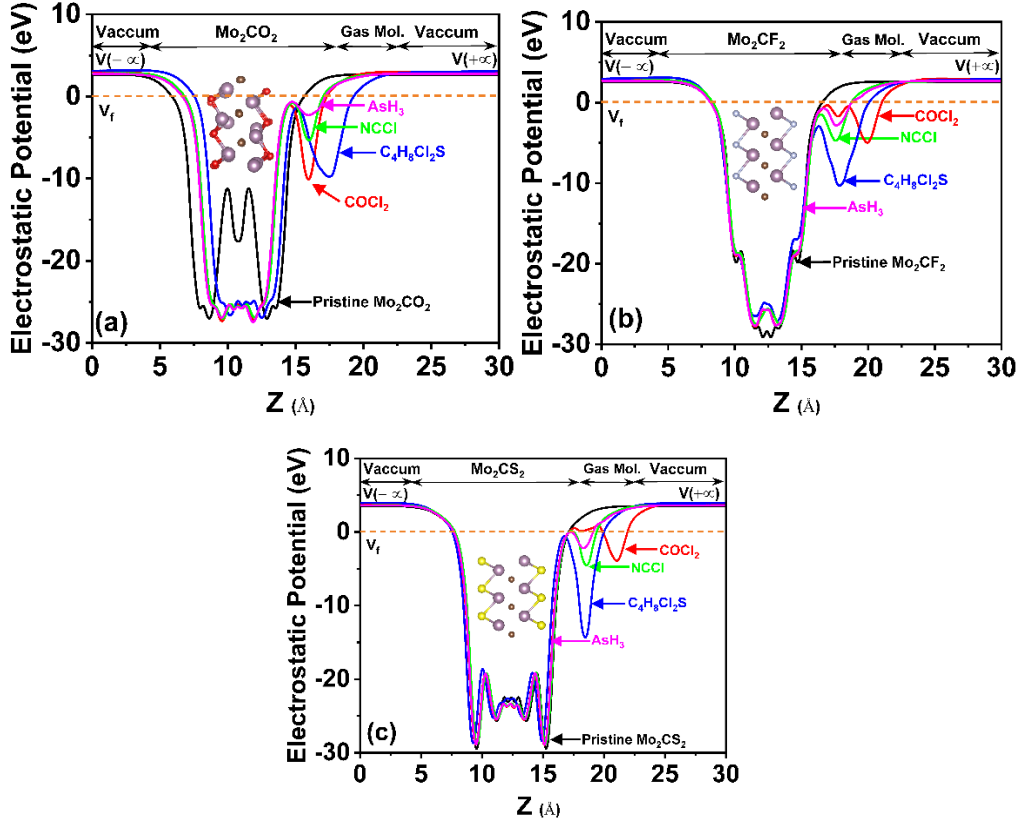

Fig.6 The planar average of the electrostatic potential of bare and CWAs adsorbed (a)$Mo_2CO_2$, (b)$Mo_2CF_2$, and (c) $Mo_2CS_2$.

Likewise, the work function (Ø) identifies the energy needed to eject an electron from the material and also describes the mobility of the electrons with variation in the resistivity. During gas adsorption reactions, the change in Ø can identify the sensitivity response of the sensor to detect the particular gas [48–50]. Following relations is used to calculate Ø

$$\emptyset = V(\infty) - \varepsilon_f \qquad (3)$$

Here, $V(\infty)$ and $\varepsilon_f$ represent the electrostatic potential at the point far from the target surface (vacuum) and Fermi level of the system, respectively.

To identify the distinct variation in resistivity, which can selectively detect the target gas molecule, we have calculated the ∅ of $Mo_2CT_x$ and $Mo_2CT_x$@CWAs systems and the results are presented in Fig. 7 (a-c). It is evident from the Fig. 7 that ∅ of bare $Mo_2CT_x$ varies upon the adsorption of CWAs, which is evidence of the variation in the conductivities.

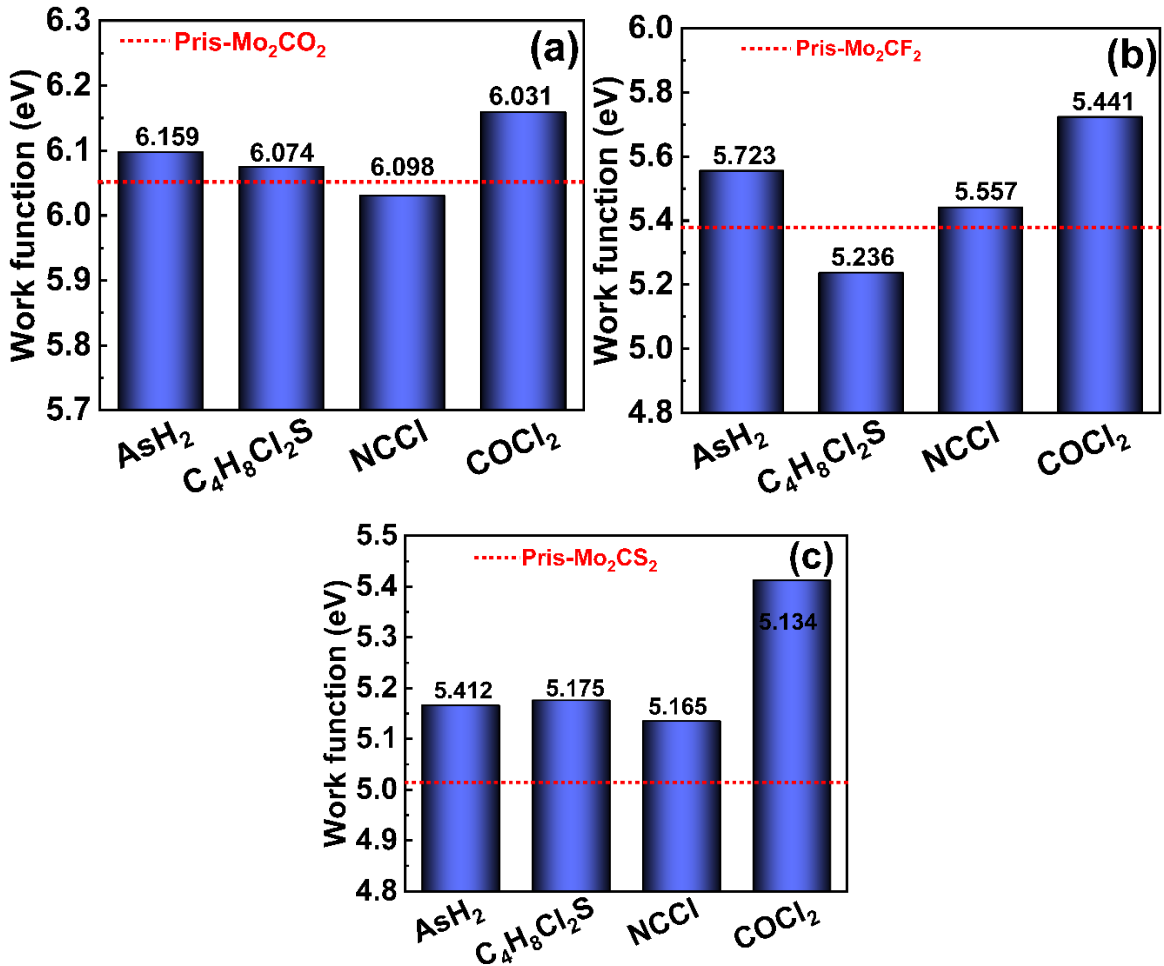

Fig.7 The calculated work-function (∅) of $Mo_2CT_x$ adsorbed with CWAs systems, (a)$Mo_2CO_2$, (b)$Mo_2CF_2$,(c) $Mo_2CS_2$. The dotted lines represent ∅ of pristine $Mo_2CT_x$.

### 3.4 Effect of external electric field

We have further investigated the sensing mechanism of CWAs on $Mo_2CT_x$ under external electric field and the resulting $E_{ads}$ values are given in Table 2. The magnitude of the electric field can be chosen according to the calculated $\omega_f$ of the $Mo_2CT_x$ and their distances with the artificial dipoles. Yeh et al., [38] expressed it as; $\exp(-\sqrt{\emptyset})$, where z is the vertical distance from $Mo_2CT_x$. Also, the distance between the surface of $Mo_2CT_x$ and the artificial dipole should not exceed $\emptyset/E_f$. Since the calculated $\emptyset$ of $Mo_2CT_x$ and $Mo_2CT_x$@CWAs are in the range of 5.01 to 6.01, with a vacuum space of 18.5 Å, we have taken $E_f$ to be + 0.2 V/Å to - 0.2 V/Å in our investigation. From Table 2, it is evident that the functional groups (-O, -S, -F) play a key role in the adsorption of CWAs, even in the presence of the external field. For $Mo_2CO_2$ there is a small change in $E_{ads}$ values of $AsH_3$ and $COCl_2$. However, the $E_{ads}$ value for $C_4H_8Cl_2S$ increases tremendously both at negative and positive electric field. On contrary, the binding of NCCl weakens in the presence of the applied field. For $Mo_2CF_2$, the applied electric field causes a significant reduction in the $E_{ads}$ values. In the absence of electric field, CWAs strongly chemisorbed over $Mo_2CF_2$, however the introduction of electric field the $E_{ads}$ values fall back to the physisorption range. Similarly, for $Mo_2CS_2$, the applied electric field plays a significant role in bringing the $E_{ads}$ from strong chemisorption to desirable physisorption range. Ground state configurations of $Mo_2CT_x$@CWAs under external eclectic fields are given in Fig. S1-5 (Supporting Information).

Table 2: Summary of electric field effects on the calculated adsorption energies ($E_{ads}$), of different CWAs on $Mo_2CT_x$.

|  | $E_{ads}$ (eV) | | | | | | | | |
| --- | --- | --- | --- | --- | --- | --- | --- | --- | --- |
|  | $Mo_2CO_2$ | | | $Mo_2CF_2$ | | | $Mo_2CS_2$ | | |
| E-Feld strength (V/Å) | +0.2 | 0 | -0.2 | +0.2 | 0 | -0.2 | +0.2 | 0 | -0.2 |
| $AsH_3$ | -0.27 | -0.29 | -0.26 | -0.25 | -4.05 | -0.23 | -0.29 | -1.98 | -0.22 |
| $C_4H_8Cl_2S$ | -0.86 | -0.38 | -0.94 | -0.04 | -3.07 | -0.17 | -1.27 | -2.54 | -1.31 |
| NCCl | -0.31 | -0.58 | -0.35 | -0.44 | -3.93 | -0.49 | -0.26 | -2.02 | -0.27 |
| $COCl_2$ | -0.40 | -0.40 | -0.37 | -0.32 | -4.12 | -0.24 | -0.29 | -1.99 | -0.25 |

### 3. 5 Adsorption mechanism in aqueous medium

For practice applications, it is important to study the interaction mechanism of CWAs with $Mo_2CT_x$ in an aqueous medium. In chemical processes where the solvent with a specific dielectric constant is treated as a continuous medium, the implicit solvation method (ISM) is often a realistic approach to calculating the energetics of solute-solvent interactions. Using ISM, we can predict the polar contribution to the free energy using the dielectric constant of the solvent. Using the VASP code, the ISM is used to calculate the effect of water on the CWAs molecules and crystal surfaces $Mo_2CT_x$. In addition, the ISM is used to calculate as their corresponding reaction barriers. CWAs and $Mo_2CX_2$ are treated quantum mechanically in the ISM. An electrochemical interface simulation is considered to be more realistic when it treats the solvent (water) as a continuum and includes the effects of ionic solution at a first-principles level. The adsorption energies of CWAs to the host $Mo_2CT_x$ are calculated as per Eq. (4) and summarized in Table 3. The optimized structure of the $Mo_2CT_x$@CWAs in an aqueous environment are presented in Fig. S5 (Supporting Information). From Table 3, it is interesting to notice that the functional groups play a crucial role in the binding of CWAs, which is very prominent in the aqueous medium. In aqueous medium, the affinity of $Mo_2CT_x$, have slightly decreased towards the studied CWAs except for $C_4H_8Cl_2S$.

$$E_{adQ} = E_{S-AAs-Aq} - E_{S-Aq} - E_{AAs} \qquad (4)$$

Table 3 Adsorption energies ($E_{ads}$) of CWAs on $Mo_2CT_x$ in the aqueous medium.

| | Adsorption Energy (eV) | | |
| --- | --- | --- | --- |
| | $E_{ads}$ | | |
| | $Mo_2CO_2$ | $Mo_2CF_2$ | $Mo_2CS_2$ |
| CWAs | | | |
| $AsH_3$ | -0.25 | -1.36 | -2.93 |
| $C_4H_8Cl_2S$ | -0.88 | -1.96 | -3.33 |
| NCCl | -0.31 | -1.55 | -2.98 |
| $COCl_2$ | -0.38 | -1.37 | -3.02 |

## 3. 6 Thermodynamic analyses

Adsorption efficiency of the selected CWA, which is presented as surface coverage was estimated by using Langmuir adsorption model and Boltzmann statistics [51]. The surface coverage $f(P,T)$ by homogeneous gas species $i$, is given by

$$f(P,T) = \frac{\exp[(\mu_i(P,T) - \varepsilon_i)/k_B T]}{1 + \exp[(\mu_i(P,T) - \varepsilon_i)/k_B T]} \quad (5)$$

where $\mu_i$ indicates the molar Gibbs free energy of the gas molecule $i$ in gaseous phase which depends on the pressure $P$ and the temperature $T$, respectively, and $\varepsilon_i$ denotes the $E_{ads}$ obtained by DFT calculation. $k_B$ is Boltzmann constant. The molar free energy of each CWA species was computed by using Gaussian09 software [52] in ωB97XD/aug-cc-pVTZ [53,54] level of theory, and the result was formulated by following equation as

$$\mu_i(P,T) = k_B T \ln\left[\frac{P}{k_B T}\left(\frac{2\pi\hbar^2}{M k_B T}\right)^{3/2}\right] + (A + BT) \quad (6)$$

where $\hbar$ and $M$ are Planck constant and molar mass of selected molecule, respectively, and the constant $A$ and $B$ are fitting parameters. These values are listed on Table 4. The first term indicates free energy of monatomic ideal gas, and the second describes excess free energy which includes thermal effect of molecular motion. In dilute conditions below 1 atm, pressure dependence is negligible in excess term.

Adsorption efficiencies in isobaric and isothermal environments are shown in Fig.8. In two scenarios, 1 ppb of constant pressure was assumed for CWAs in order to examine the operating temperature range, and 300 K of constant temperature was set to evaluate the detection capabilities of CWA gases at dilute concentrations. Our results show $Mo_2CO_2$ is not suitable for detecting CWAs as a chemical sensor because of non-spontaneous adsorption under practical conditions. In contrast, $Mo_2CF_2$ and $Mo_2CS_2$ exhibit complete coverages in ambient and warmer conditions, up to 500 K and 400 K for $Mo_2CF_2$ and $Mo_2CS_2$, respectively. Furthermore, they are maintained down to parts per quadrillions (ppq) of gas pressure at 300 K. Due to greater adsorption energies, $Mo_2CF_2$ is anticipated to exhibit superior CWA adsorption performances than $Mo_2CS_2$ at elevated temperatures, as depicted in Fig.8a. This result denotes that $Mo_2CF_2$ can be a viable CWAs chemical sensor with excellent detecting capabilities for rarified toxic gases and thermal stabilities.

Table 4. Fitting parameters in Eq. (6) and their coefficient of determination ($R^2$) values for each CWA species.

| Toxic CWAs | $A$ (eV) | $B$ (meV/K) | $R^2$ |
|---|---|---|---|
| $AsH_3$ | 0.07357 | -0.81062 | 0.9979 |
| $C_4H_8Cl_2S$ | 0.41946 | -3.28274 | 0.9950 |
| NCCl | 0.08199 | -0.89061 | 0.9978 |
| $COCl_2$ | 0.12723 | -1.39444 | 0.9977 |

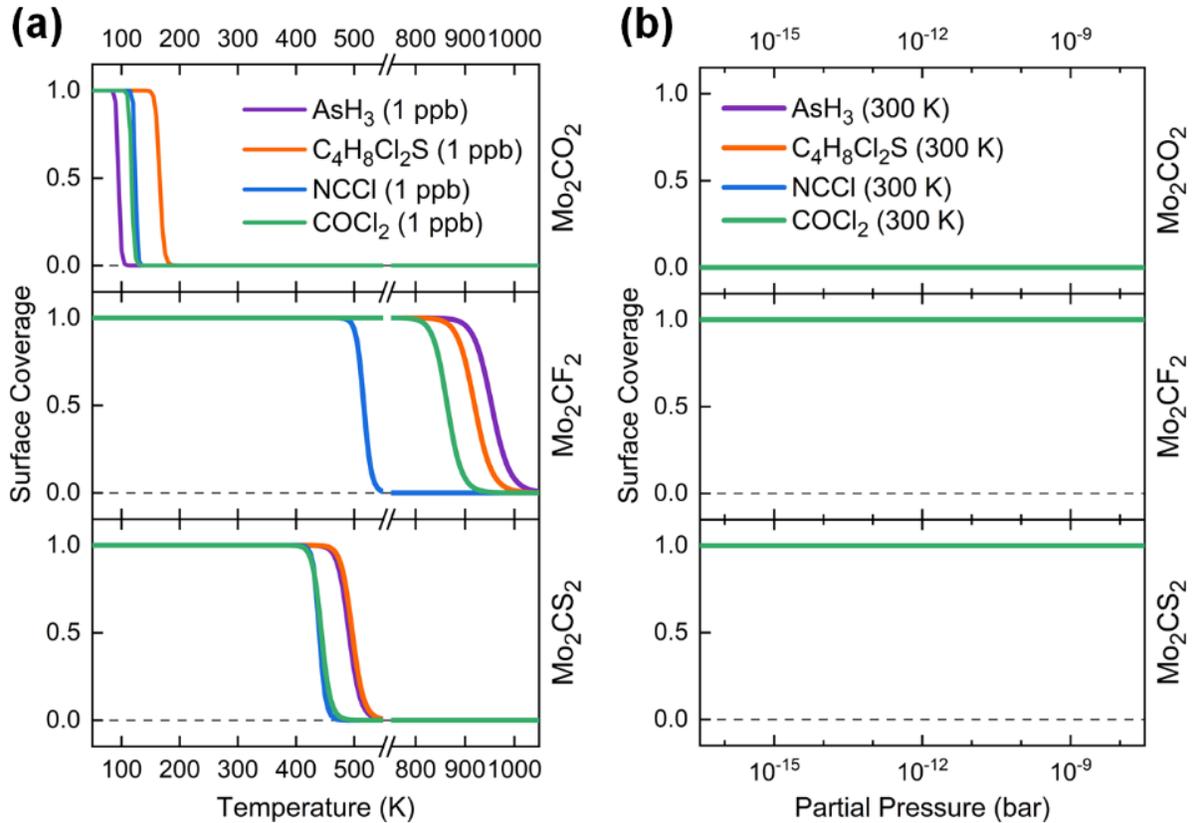

**Fig. 8** Surface coverage of CWA on $Mo_2CT_x$ surfaces expressed as (a) a function of temperature in 1 ppb of constant partial pressure and (b) a function of partial pressure in 300 K of temperature. All surface coverage profiles for each gas in (b) are overlapped. $Mo_2CO_2$ substrate is unable to capture target gases in ambient condition because of their non-spontaneous adsorption. On the other hand, $Mo_2CF_2$ and $Mo_2CS_2$ show excellent adsorption efficiencies for all target gases even in rarified conditions about $10^{-15}$ bar because of strong adsorption energies. It denotes all contacted target molecules on substrate surface can be captured until the surface are fully covered.

## 4. Summary:

We used spin-polarized DFT calculations to study the structural, electronic and sensing properties of selected chemical warfare agents (CWAs) by molybdenum carbide based MXenes ($Mo_2CT_x$; $T_x$= O, F, S). The considered CWAs included arsine ($AsH_3$), mustard gas ($C_4H_8Cl_2S$), cyanogen chloride (NCCl), and phosgene ($COCl_2$), which are extremely toxic agents and had been used as chemical weapons in the past. Van der Waals corrected energy calculations revealed strong adsorption energies ($E_{ads}$) of CWAs, in the range of -2.33 to -4.05 eV, on $Mo_2CF_2$, and $Mo_2CS_2$ monolayers However, $Mo_2CO_2$ adsorbed CWAs moderately with $E_{ads}$ values of -0.29 to -0.58 eV, which indicated reversible sensing mechanism. Distinct variations in the electronic structures, electrostatic potentials and work functions of $Mo_2CT_x$ were observed upon the adsorption of CWAs. This indicated an effective sensing mechanism of the $Mo_2CT_x$ monolayers towards the studied CWAs. In addition to gas phase, we performed solvation calculations to study the sensing mechanism of $Mo_2CT_x$ in aqueous medium. Finally, the effect of pressure and temperatures on the sensing properties of $Mo_2CT_x$ monolayers were considered through statistical thermodynamic analysis. Our findings clearly showed promising sensing characteristics of $Mo_2CT_x$ monolayers towards selected CWAs, and thus pave the way to an innovative class of reusable nanosensors for defence applications.

## Acknowledgments

PP is indebted to the CENCON for financial support. This research was supported by the Fundamental Fund of Khon Kaen University (project number 161739). The research has received funding support from the National Science, Research and Innovation Fund (NSRF). TH acknowledges the computational resources provided at the NCI National Facility systems at the Australian National University through the National Computational Merit Allocation Scheme supported by the Australian Government. Computational resources provided by the Pawsey Supercomputing Centre are also acknowledged.